\documentclass[aps,prl,nofootinbib,tightenlines,notitlepage, twocolumn,floatfix,superscriptaddress,showkeys]{revtex4-1}
    
\usepackage[utf8]{inputenc}          
\usepackage{graphicx}         
\usepackage[usenames,dvipsnames]{xcolor}
\usepackage{array,dcolumn} 
\usepackage{soul}

\usepackage{enumitem}

\usepackage{amsmath,amssymb,amsfonts,slashed} 

\usepackage[linktocpage,breaklinks]{hyperref}

\usepackage{txfonts}
\usepackage{bm}
\usepackage{stmaryrd}
\usepackage[utf8]{inputenc}

\usepackage{epsfig}
\usepackage{epstopdf}

\usepackage{natbib}
\usepackage{cleveref}

\definecolor{mred}{RGB}{127,0,25}
\definecolor{mdgr}{RGB}{51,51,51}
\definecolor{mag}{RGB}{211, 54, 130}
\definecolor{verm}{RGB}{164, 25, 0}

\hypersetup{colorlinks=true,
            citecolor=NavyBlue,
            linkcolor=NavyBlue,
            urlcolor=NavyBlue}

\usepackage{siunitx}                                  
\DeclareSIUnit{\fm}{\femto\metre}                     

\begin{document}

\title{The slope, the hill, the drop, and the swoosh: \\
Learning about the nuclear matter equation of state from the binary Love relations}
\date{\today}

\author{Hung Tan}
\affiliation{Illinois Center for Advanced Studies of the Universe, Department of Physics, University of Illinois at Urbana-Champaign, Urbana, IL 61801, USA}
\author{Veronica Dexheimer}
\affiliation{Department of Physics, Kent State University, Kent, OH 44243 USA}
\author{Jacquelyn Noronha-Hostler}
\affiliation{Illinois Center for Advanced Studies of the Universe, Department of Physics, University of Illinois at Urbana-Champaign, Urbana, IL 61801, USA}
\author{Nicol\'as Yunes}
\affiliation{Illinois Center for Advanced Studies of the Universe, Department of Physics, University of Illinois at Urbana-Champaign, Urbana, IL 61801, USA}

\begin{abstract}

Analyses that connect astrophysical observations of neutron stars with nuclear matter properties sometimes rely on equation-of-state insensitive relations. We show that the slope of the binary Love relations (i.e.~between the tidal deformabilities of binary neutron stars) encodes the rate of change of the nuclear matter speed of sound below three times nuclear saturation density. Twin stars lead to relations that present a signature ``hill,'' ``drop,'' and ``swoosh'' due to the second (mass-radius) stable branch, requiring a new description of the binary love relations.

\end{abstract}

\maketitle

\vspace{0.2cm}
\noindent \textit{Introduction}~--~
Neutron stars (NSs) contain the highest possible densities known to humanity within their core.  The maximum baryon density  can reach up to 10 times nuclear saturation density ($\rho_{sat} \equiv m_N n_{\rm sat} \approx 2.7 \times 10^{14}~{\rm{g}} \; {\rm{cm}}^{-3}$, with $m_N$ the nucleon mass and $n_{\rm sat} \equiv 0.16~{\rm{fm}}^{-3}$ the baryon number density at saturation). As a consequence, their cores may contain baryon resonances and hyperons, as well as deconfined quark matter. It is not yet clear if the appearance of these new degrees of freedom  would be related to smooth cross-overs or $n^{th}$-order phase transitions. In this letter, we explore specific features in the equation of state (EoS) related to phase transitions that lead to observable features in the binary Love relations (BLR) of NSs~\cite{Yagi:2013awa,Yagi:2013bca,Yagi:2015pkc}.

For $1^{st}$-order phase transitions, additional stable branches may appear in mass-radius sequences giving rise to \textit{mass twins} \cite{Benic:2014jia,Christian:2020xwz,Han:2020adu,Pang:2020ilf,Jakobus:2020nxw,Tan:2021ahl,Alford:2013aca,Alford:2015dpa,Alford:2015gna,Ranea-Sandoval:2015ldr,Han:2018mtj,Chatziioannou:2019yko,Blaschke:2020vuy,Blaschke:2020vuy,Pang:2020ilf,Dexheimer:2014pea,Mishustin:2002xe,Jakobus:2020nxw,Alford:2017qgh,Zacchi:2015oma,Alvarez-Castillo:2018pve,Li:2019fqe,Wang:2019npj,Fadafa:2019euu,Blaschke:2020qrs,Christian:2019qer,Benitez:2020fup,Sharifi:2021ead,Deloudis:2021agp,Li:2021sxb,Christian:2021uhd,Goncalves:2021pmr,Espino:2021adh}, ie.~NSs with nearly the same mass but different radii.
For cross-overs, large ``bumps'' in the speed of sound ($c_s^2 \equiv dp/d\epsilon$ where $p$ is the pressure and $\epsilon$ the energy density) can appear
\cite{Dexheimer:2014pea,Dutra:2015hxa,McLerran:2018hbz,Jakobus:2020nxw,Alford:2017qgh,Zacchi:2015oma,Alvarez-Castillo:2018pve,Li:2019fqe,Wang:2019npj,Fadafa:2019euu,Xia:2019xax,Yazdizadeh:2019ivy,Shahrbaf:2019vtf,Zacchi:2019ayh,Zhao:2020dvu,Lopes:2020rqn,Blaschke:2020qrs,Duarte:2020xsp,Rho:2020eqo,Marczenko:2020wlc,Minamikawa:2020jfj,Hippert:2021gfs,Pisarski:2021aoz,Sen:2020qcd,Stone:2021ngh,Ferreira:2020kvu,Kapusta:2021ney,Somasundaram:2021ljr,Kojo:2020krb,Mukherjee:2017jzi,Dexheimer:2017nse,Li:2020dst,Jimenez:2021wil,Jin:2021xkz,Lee:2021hrw,Raduta:2021coc}.  A fast increase in $c_s^2$ points to a rapid change in the characteristics of strongly interacting matter, due to 
(i) appearance of strangeness or more massive degrees of freedom,
(ii) change in effective degrees of freedom,
(iii) restoration or breaking of symmetries, and
(iv) strengthening or weakening of different strong interactions. 
Reason (i) could be related to stiff EoSs suddenly softening due to the appearance of hyperons, resonances,  or strange quarks (e.g. Fig.~2 in \cite{Tan:2021ahl} and~\cite{Jakobus:2020nxw,Dexheimer:2014pea,Dutra:2012mb,McLerran:2018hbz,Alford:2017qgh,Zacchi:2019ayh,Zhao:2020dvu,Minamikawa:2020jfj,Hippert:2021gfs,Pisarski:2021aoz,Sen:2020qcd,Stone:2021ngh,Kapusta:2021ney,Somasundaram:2021ljr,Dexheimer:2009hi,Dexheimer:2008ax,Guichon:1995ue,Malfatti:2020onm,Bedaque:2014sqa,Annala:2019puf,Dexheimer:2020rlp,Ferreira:2020kvu,Jakobus:2020nxw,Dexheimer:2014pea,Dutra:2015hxa,McLerran:2018hbz,Alford:2017qgh,Zacchi:2019ayh,Zhao:2020dvu,Minamikawa:2020jfj,Hippert:2021gfs,Pisarski:2021aoz,Sen:2020qcd,Stone:2019blq,Kapusta:2021ney,Somasundaram:2021ljr,Bedaque:2014sqa,Malfatti:2020onm,Guichon:1995ue,Dexheimer:2008ax,Dexheimer:2009hi,Gusakov:2014ota,dePaoli:2012cz,Sen:2021bms,Tu:2021sxx}.
Reason (ii) could be related to first-order phase transitions \cite{Blaschke:2018pva,Montana:2018bkb,Kojo:2019raj,Malfatti:2019tpg,Shahrbaf:2019vtf,Marczenko:2020wlc,Jokela:2020piw,Somasundaram:2021ljr} or an “unphase transition”, as predicted by the quarkionic model \cite{Jeong:2019lhv,McLerran:2018hbz,Duarte:2020xsp,Zhao:2020dvu,Sen:2020qcd,Sen:2020peq} or crossover transitions/Gibbs constructions to/with (stiff) quark matter \cite{Motornenko:2019arp,Baym:2019iky,Li:2018ltg,Xia:2019xax,Yazdizadeh:2019ivy,Minamikawa:2020jfj,Kapusta:2021ney,Kumar:2021hzo}.
Reason (iii) could be associated with chiral symmetry (predicted by QCD to be restored at large densities) \cite{Mukherjee:2017jzi,Motornenko:2019arp,Marczenko:2021uaj,Dexheimer:2008cv} or symmetries related to different pairing schemes \cite{Alford:2017qgh,Hippert:2021gfs}. 
Reason (iv) could be related to the behavior of different meson condensates. \cite{Pisarski:2021aoz}. 

Gravitational wave (GW) observations may reveal the complexity of matter at high densities through inferences on the masses and tidal deformabilities of NSs in coalescing binaries. 
In a binary, the gravitational field of one star perturbs the field of the other and vice-versa, speeding up their inspiral. 
Since GWs are generated by accelerations, this change in the inspiral rate is encoded in the GWs emitted,  which carry information about the NS tidal deformabilities $\Lambda_{1,2}$.
Thus, a sufficiently loud signal allows for a measurement of a certain combination of $\Lambda_{1,2}$, the chirp tidal deformability $\tilde{\Lambda}$. One method to extract $\Lambda_{1,2}$ from $\tilde{\Lambda}$ is through EoS-insensitive BLR~\cite{Yagi:2013awa,Yagi:2013bca,Yagi:2015pkc,Chatziioannou:2018vzf,LIGOScientific:2018cki}. Once $\Lambda_1$ and $\Lambda_2$ have been inferred, one can then use a second EoS-insensitive ``Love-C'' relation between $\Lambda_{1,2}$ and the compactness ${\mathcal{C}}_{1,2} \equiv M_{1,2}/R_{1,2}$ (stellar mass over radius) to infer the individual compactnesses. From these, and separate inferences about the component masses, encoded in other parts of the GW, one can  infer the individual radii of the NSs~\cite{Yagi:2015pkc,Chatziioannou:2018vzf,LIGOScientific:2018cki}. 
These EoS-insensitive relations were originally derived using a vast array of EoSs that do not present features such as bumps in $c_s^2$, nor do they allow for mass twins. Could it be that structure in the  $c_s^2$ imprints onto the EoS-insensitive relations, thus allowing for the possibility of discovering their existence with GW observations of inspiraling NSs?
This is the question we tackle here. 

\vspace{0.2cm}
\noindent \textit{Methodology}~--~ 
We follow the formalism in~\cite{Tan:2021ahl} to model the EoS, which consists of three parts: a ``crust" (loosely defined as the low density EoS $n \lesssim 1.5~n_{\rm sat}$), structure functions, and transition functions. For the crust, we use the SLy EoS \cite{Chabanat:1997un,Chabanat:1997un,Douchin:2000kad,Douchin:2000kx}. Beyond that, a bump and/or plateau is introduced in $c_s$ to mimic different-order phase transitions. Transition functions are used to connect different parts of the EoS smoothly~\cite{Tan:2020ics,Tan:2021ahl}.

We then follow~\cite{Yagi:2013awa,Yagi:2013bca} to calculate the mass, radius, and tidal deformabilities of non-rotating NSs in a binary system during the late quasi-circular inspiral. 
At zeroth-order in perturbation theory, the Einstein equations reduce to the  Tolman-Oppenheimer-Volkov (TOV) equations, which we solve numerically to find the radius and the mass for a given central density. At first-order in perturbation theory, we solve the linearized Einstein equations to find the (electric-type, quadrupole) tidal deformability, defined as minus the ratio of the induced quadrupole moment to the external quadrupole tidal field. We then non-dimensionalize this quantity to compute the dimensionless tidal deformability $\Lambda$ at a given central density. Considering a sequence of central densities for a given EoS, we can then construct a mass-radius curve, a Love-C curve, and for a fixed mass ratio, the BLRs $\Lambda_a(\Lambda_s)$, defined with the symmetric and anti-symmetric tidal deformabilities $\Lambda_{s,a} = (\Lambda_1 \pm \Lambda_2)/2$~\cite{Yagi:2013awa,Yagi:2013bca}.

\begin{figure*}[ht]
\centering
\begin{tabular}{c c c c}
\includegraphics[width=0.325\linewidth]{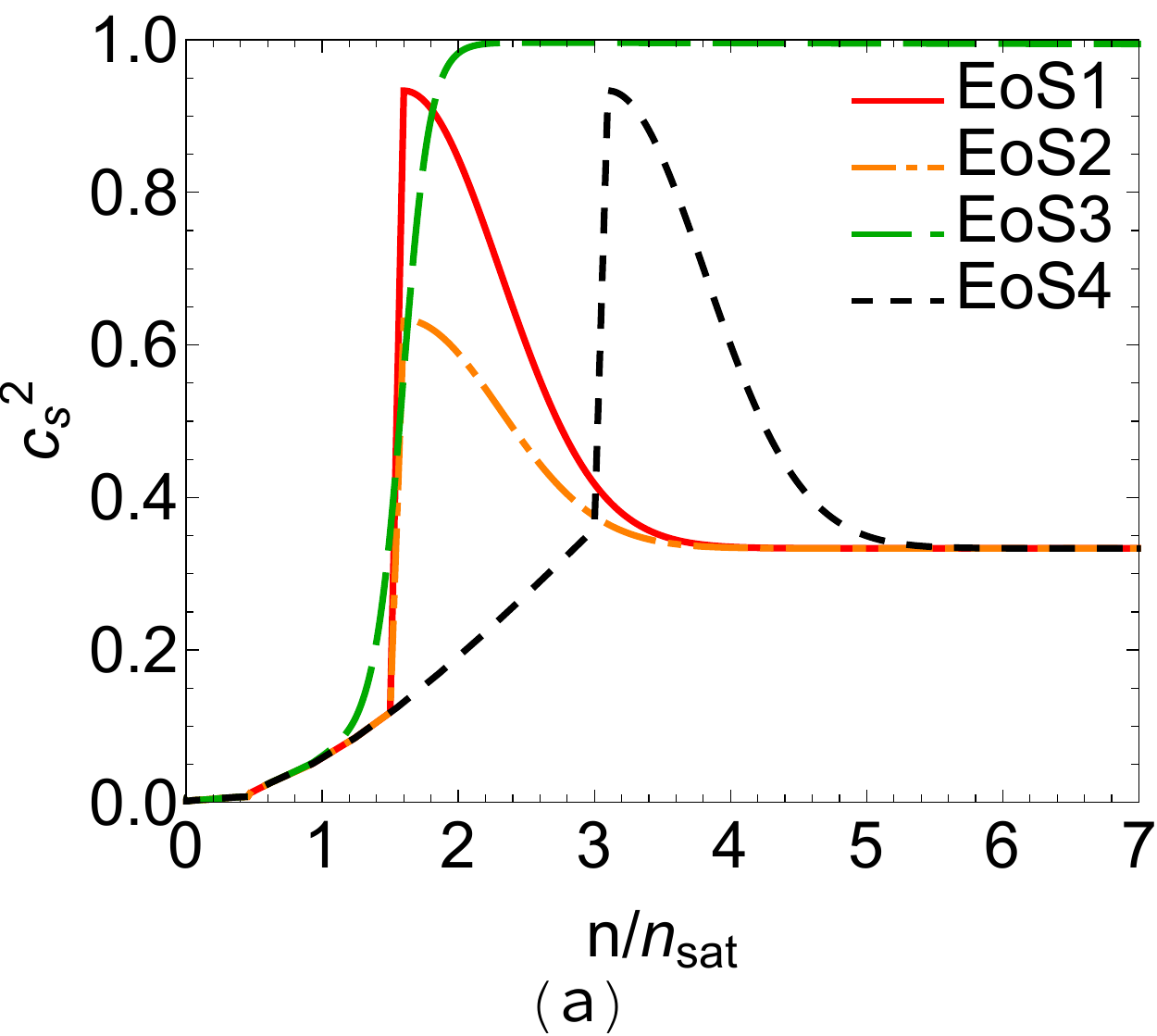} &
\includegraphics[width=0.325\linewidth]{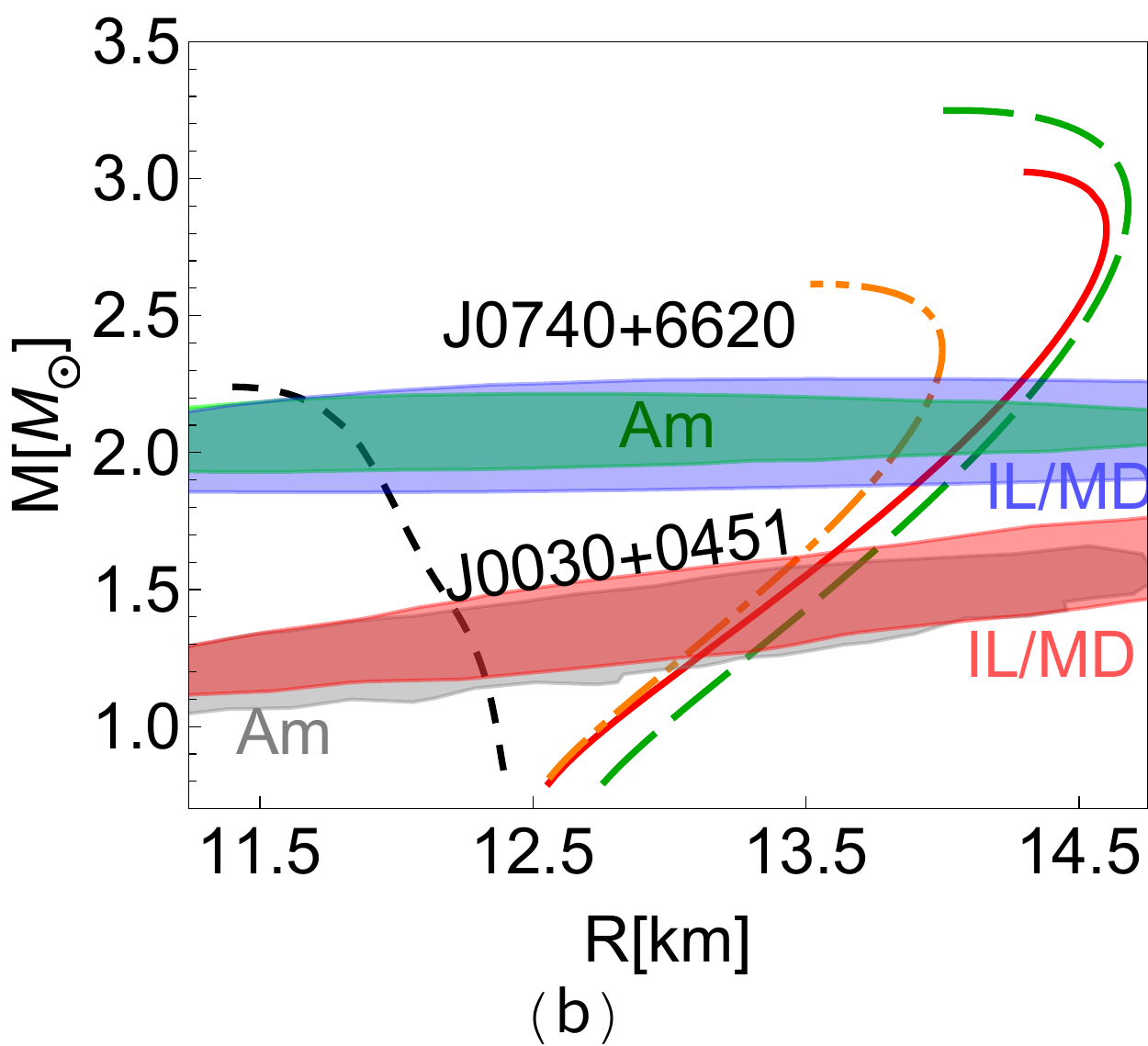} &
\includegraphics[width=0.325\linewidth]{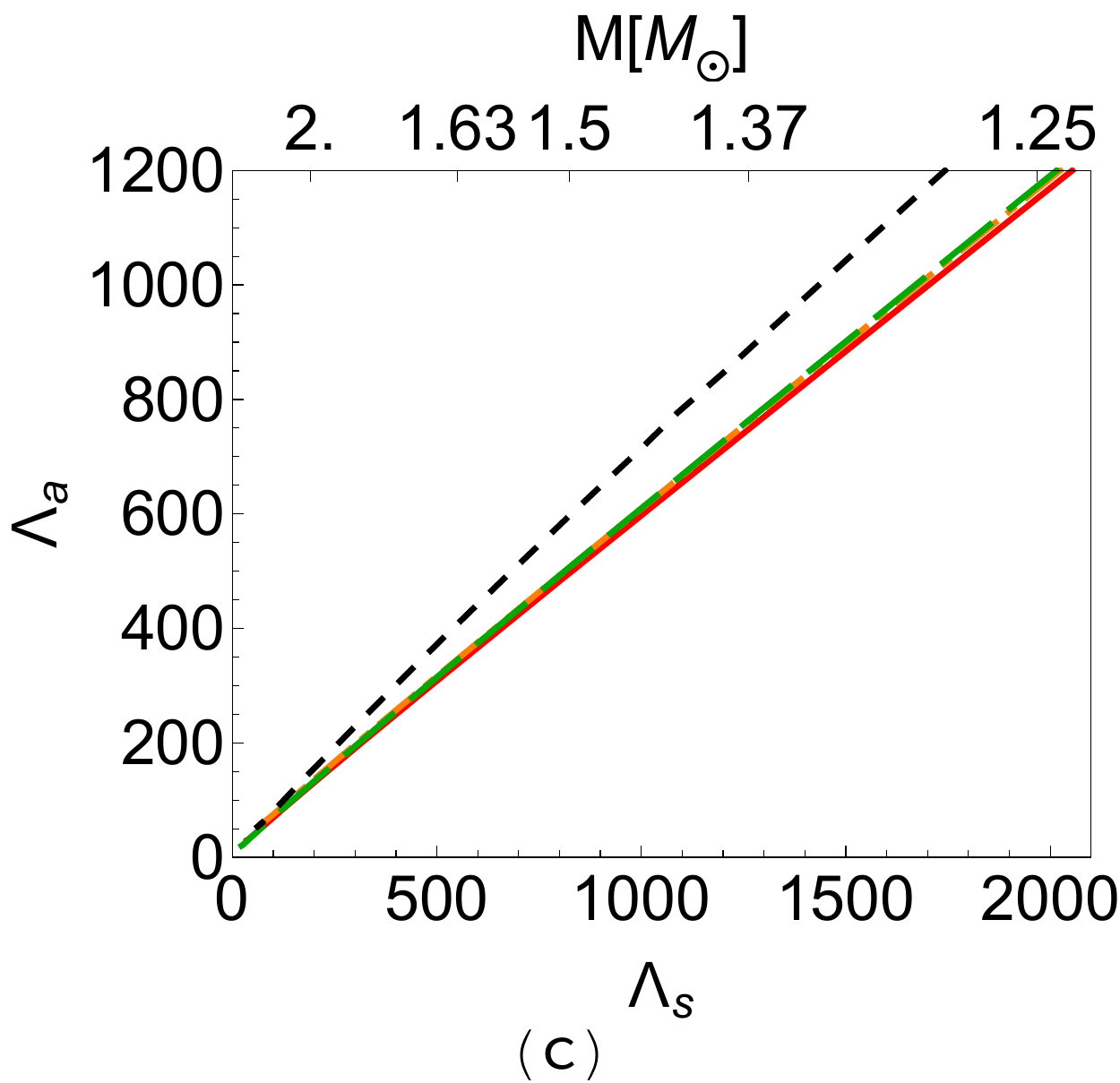} &
\end{tabular}
\caption{(Color online)  $c_s^2$ as a function of $n/n_{\rm sat}$ (a), mass-radius curves (b) and BLRs for a mass ratio of $q = 0.75$ (c) for 4 EoS models. 
Structure in the EoSs near saturation bends the mass-radius curves to larger radii, while still being consistent with the two NICER observations at 90\% confidence (Amsterdam or Illinois/Maryland). 
} 
\label{fig:slope}
\end{figure*}

\vspace{0.2cm}
\noindent \textit{Mass and radius without $1^{st}$-order phase transitions}~--~
Let us begin by considering NSs with EoSs that present structure in $c_s^2$ with higher (than first)-order phase transitions. 
Panel (a) of Fig.~\ref{fig:slope} shows $c_s^2$ for 4 EoS models. The first three present a steep rise in $c_s^2$ that starts around $n \lesssim 1.5~n_{\rm sat}$, followed either by a plateau at the causal limit or a bump that returns to the conformal limit. EoS 4 presents a bump similar to that of EoS 1, but at a larger number density ($n \sim 3~n_{\rm sat}$). 

The corresponding mass-radius curves are shown in panel (b) of Fig.~\ref{fig:slope}. EoSs that 
have a steep rise in $c_s^2$ at densities of $n \gtrsim 3~n_{\rm sat}$ lead to heavy NSs where the mass-radius curves bends towards the left (i.e. $dM/dR<0$ for EoS 4). In contrast, EoSs with a steep rise in $c_s^2$ at lower densities produce mass-radius sequences that bend to the right (i.e. $dM/dR>0$ for EoS 1--3). Why is this? A sharp rise in $c_s^2$ implies a faster increase of pressure with baryon density. If the sudden rise happens at low density, i.e.~somewhere relatively close to the stellar surface, the pressure inside the NS is larger than it would be otherwise, which tends to push the radius of the NS to larger values. Such an increase in $R$ changes the slope, $dM/dR$, from very large in absolute value and negative (EoS 4) to not as large and positive (EoSs 1--3).

Rises at  densities below $n < 1.5~n_{\rm sat}$  may also produce $dM/dR>0$ sequences. These, however, are much more constrained because such rises produce EoSs with large symmetry energy slope $L$ and fluffy stars, often outside constraints of the LVC collaboration and NICER. Thus, any rise in $c_s^2$ at low densities must be significantly smaller and accompanied by a previously mentioned EoS softening to fit within known constraints, leading to a smaller $dM/dR>0$ \cite{Dexheimer:2018dhb}. One method of circumventing this is to place a significant first-order phase transition around $n_{\rm sat}$, followed by a steep rise.

\vspace{0.2cm}
\noindent \textit{The slope}~--~Using the same EoSs without first-order phase transitions, we can compute the dimensionless tidal deformabilities of two stars in a binary, and for a fixed mass ratio we can plot the BLR, shown in panel (c) of Fig.~\ref{fig:slope}. These are linear relations, as expected from~\cite{Yagi:2013awa,Yagi:2013bca}, but the slope is different when using EoSs with bump structure at low densities. The results are shown for a fixed mass ratio: $q \equiv M_1/M_2 = 0.75$, but we have verified numerically that the qualitative behavior exists at all other mass ratios. 
The slope is independent of the bump height (EoS 1 vs.~2), or whether the structure itself is a bump or a plateau (EoS 1 vs.~3). The slope, however, does encode the number density at which the structure first appears (as shown by the change in slope between EoS 1,2,3 vs.~4). 

Why does the slope of the BLRs depend on where the structure in $c_s^2$ is first introduced? One can construct an analytic expression for the tidal deformability of the form $\Lambda = \Lambda(\mathcal{C},y)$ (e.g.~\cite{Yagi:2013awa,Yagi:2013bca}), which depends both on the compactness and a dimensionless quantity $y \equiv R h_2'(R)/h_2(R)$, where $h_2$ is related to the $(t,t)$ component of the metric. This quantity must be computed after solving the linearized Einstein equations for $h_2$, whose value at the stellar surface depends on integration constants that must be chosen to ensure the interior and exterior solutions are continuous at the surface.
While y depends on both the central density and EoS, one can always fit it to a polynomial in compactness $y = 1 + \sum_{n=1} a_n \, {\mathcal{C}}^n$, with EoS-dependent coefficients $a_n$.

Evaluating this expression for star 1 and star 2 in a binary, we can then compute the ratio $\Lambda_a/\Lambda_s$. The resulting expression is un-illuminating, but to understand the change in slope of the BLR, it suffices to consider an expansion in ${\cal{C}}_{1,2} \ll 1$:
\begin{align}
    \frac{\Lambda_a}{\Lambda_s} &= 
    \frac{\mathcal{C}_2^5-\mathcal{C}_1^5}{\mathcal{C}_2^5+\mathcal{C}_1^5} 
    +
    \frac{5(3+a_1)}{2} \frac{\mathcal{C}_1^5 \mathcal{C}_2^5 \left(\mathcal{C}_2 - \mathcal{C}_1 \right)}{(\mathcal{C}_1^5+\mathcal{C}_2^5)^2}
    + {\cal{O}}(\mathcal{C}_{1,2}^3)\,,
\label{eq:Lratio}      
\end{align}
which is consistent with the fact that $\Lambda_{1,2} \sim {\mathcal{C}}_{1,2}^{-5}$ when ${\cal{C}}_{1,2} \ll 1$, as shown in~\cite{Yagi:2013awa,Yagi:2013bca,Yagi:2015pkc}. We see that, to leading order in ${\cal{C}}$,  $\Lambda_a/\Lambda_s$ is a constant related to the difference in compactness of the stars in the binary. A linear regression of the data used in panel (c) of Fig.~\ref{fig:slope} reveals that $(\Lambda_a/\Lambda_s)_{\rm{lin.reg}} \approx 0.57$ for EoSs 1-3 and $(\Lambda_a/\Lambda_s)_{\rm{lin.reg}} \approx 0.65$ for EoS 4, while our approximation above to leading order in compactness gives $(\Lambda_a/\Lambda_s)_{\rm{Newt}} \approx 0.56$ for EoS 1-3 and $(\Lambda_a/\Lambda_s)_{\rm{Newt}} \approx 0.64$ for EoS 4. 
The approximation in Eq.~\eqref{eq:Lratio} matches the fully-numerical BLRs to better than $\lesssim 15\%$ for $\Lambda_s > 300$. 

We have shown that the BLRs are given approximately by the difference in compactnesses, but we have yet to show how this relates to the slope of the mass-radius curve. For masses sufficiently smaller than the maximum mass, we can approximate the radius as a Taylor expansion $R_2 = R_1 + (dR/dM)_1 (M_2 - M_1) + {\cal{O}}[(M_2-M_1)^2]$, where here $M_2>M_1$, but $R_2$ can be either larger or smaller than $R_1$ depending on the sign of the inverse slope $(dR/dM)_1$. Since $dM/dR$ is large for NSs, then $dR/dM$ is small (and not of ${\cal{O}}(1/{\mathcal{C}})$), and so we can write
\begin{equation}
{\mathcal{C}}_2 = \frac{{\mathcal{C}}_1}{q} \left[1 - {\mathcal{C}}_1 \left(\frac{dR}{dM}\right)_1 \left(\frac{1}{q}-1\right) + {\cal{O}}({\mathcal{C}}_1^2)\right]\,.
\label{eq:Taylor-C}
\end{equation} 
The leading-order term in the slope of the BLRs is then
\begin{align}
    \frac{\Lambda_a}{\Lambda_s} &\approx  \frac{\mathcal{C}_2^5-\mathcal{C}_1^5}{\mathcal{C}_2^5+\mathcal{C}_1^5} = \frac{1 - q^5}{1+q^5} - 10 C_1 \left(\frac{dR}{dM}\right)_1 \frac{q^4 (q-1)}{(q^5 + 1)^2} + {\cal{O}}(C_1^2)\,.
    \label{eq:LoveofSlope}
\end{align}
The slope of the BLRs clearly depends on the slope of the mass-radius curve. When $dM/dR$ is very large (i.e. EoS 4), then $dR/dM$ is very small, and the second term in Eq.~\eqref{eq:LoveofSlope} is negligible. When $dM/dR$ is not as large (EoSs 1--3), then $dR/dM$ is not so small, and it corrects the BLR slope by making the slope smaller, as shown in panel (c) of Fig.~1. 

What physical mechanisms  produce such changes in a realistic EoS?  Several scenarios have already been discussed in the introduction. In particular, we have tested the inclusion of a higher-order vector $\omega^4$ interaction in the Chiral Mean Field (CMF) model~\cite{Dexheimer:2020rlp}, which changes $c_s^2$ by introducing a rise at $n_B\sim 2~n_{\rm sat}$ and, in turn, changes the slope of the mass-radius curve at small to intermediate masses (below $1.5 M_\odot$). The inclusion of another vector interaction with iso-vector components $\omega$--$\rho$ softens the EoS around $n_B\sim~n_{\rm sat}$, forcing $c_s^2$ to become small, followed by a steep rise. This decreases the radii of low-mass stars,
tilting the mass-radius curve to the right and decreasing the slope of the BLRs (verified using the NL3 model~\cite{Pais:2016xiu,Dexheimer:2018dhb}). A similar situation arises for pure (self-bound) quarks stars \cite{Witten:1984rs}, which also exhibit a dramatic $dM/dR<0$ behavior \cite{Haensel:1986qb}. While there are subtleties in the understanding of the physical mechanism that produces $dM/dR<0$, in all cases it leads to a larger slope in $\Lambda_a(\Lambda_s)$.

\begin{figure*}
\centering
\begin{tabular}{c c c c}
\includegraphics[width=0.25\linewidth]{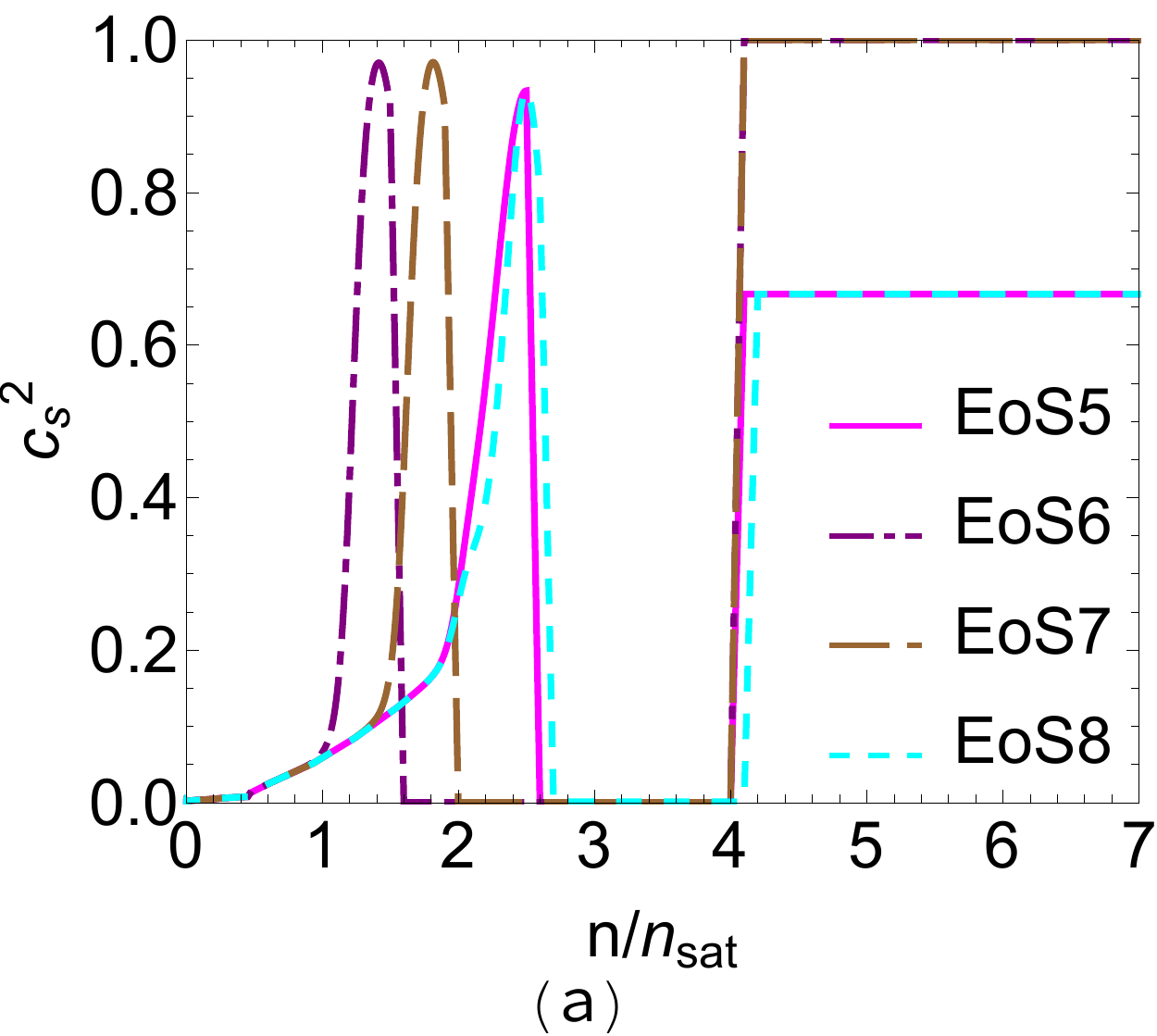} &
\includegraphics[width=0.25\linewidth]{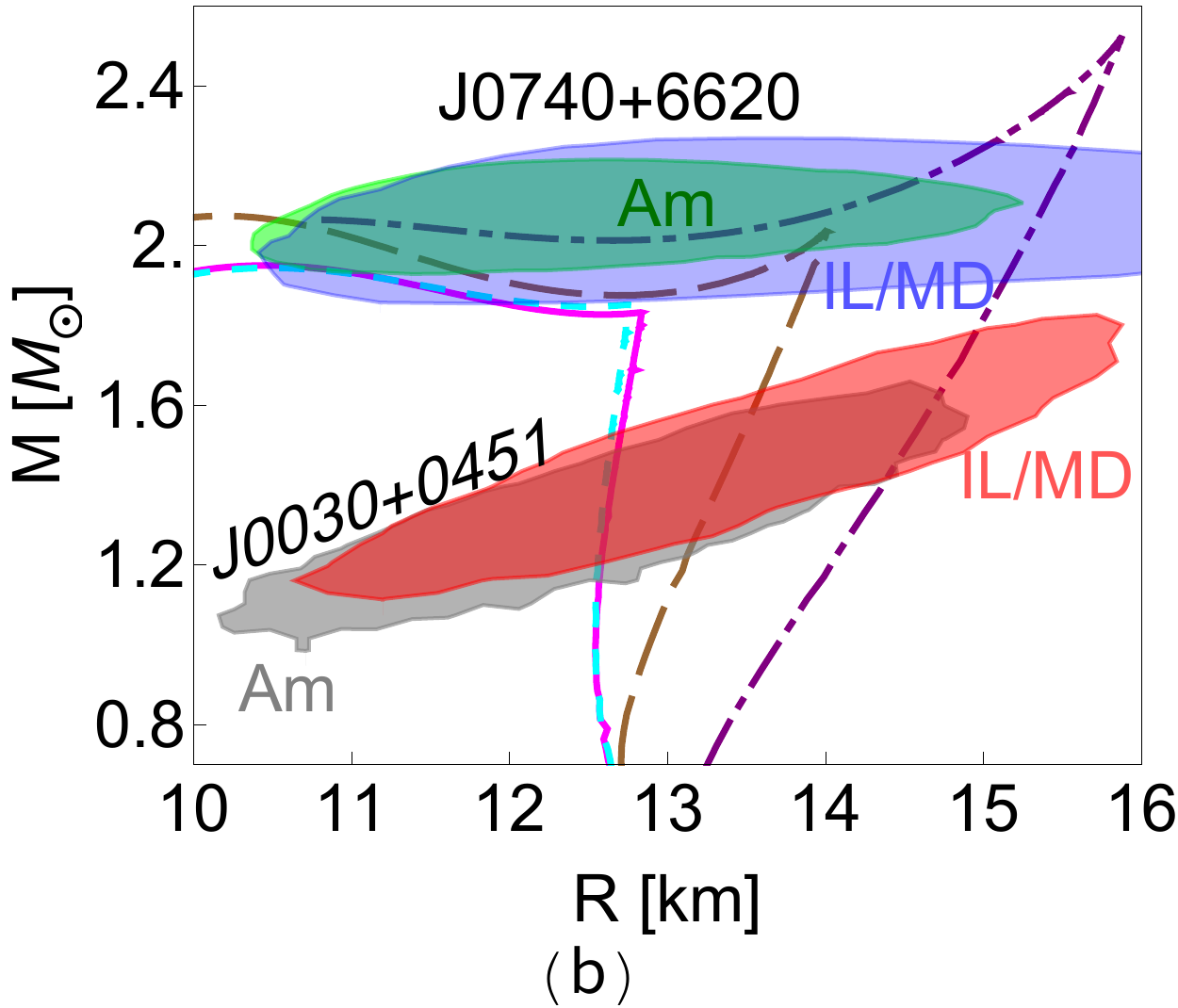} &
\includegraphics[width=0.25\linewidth]{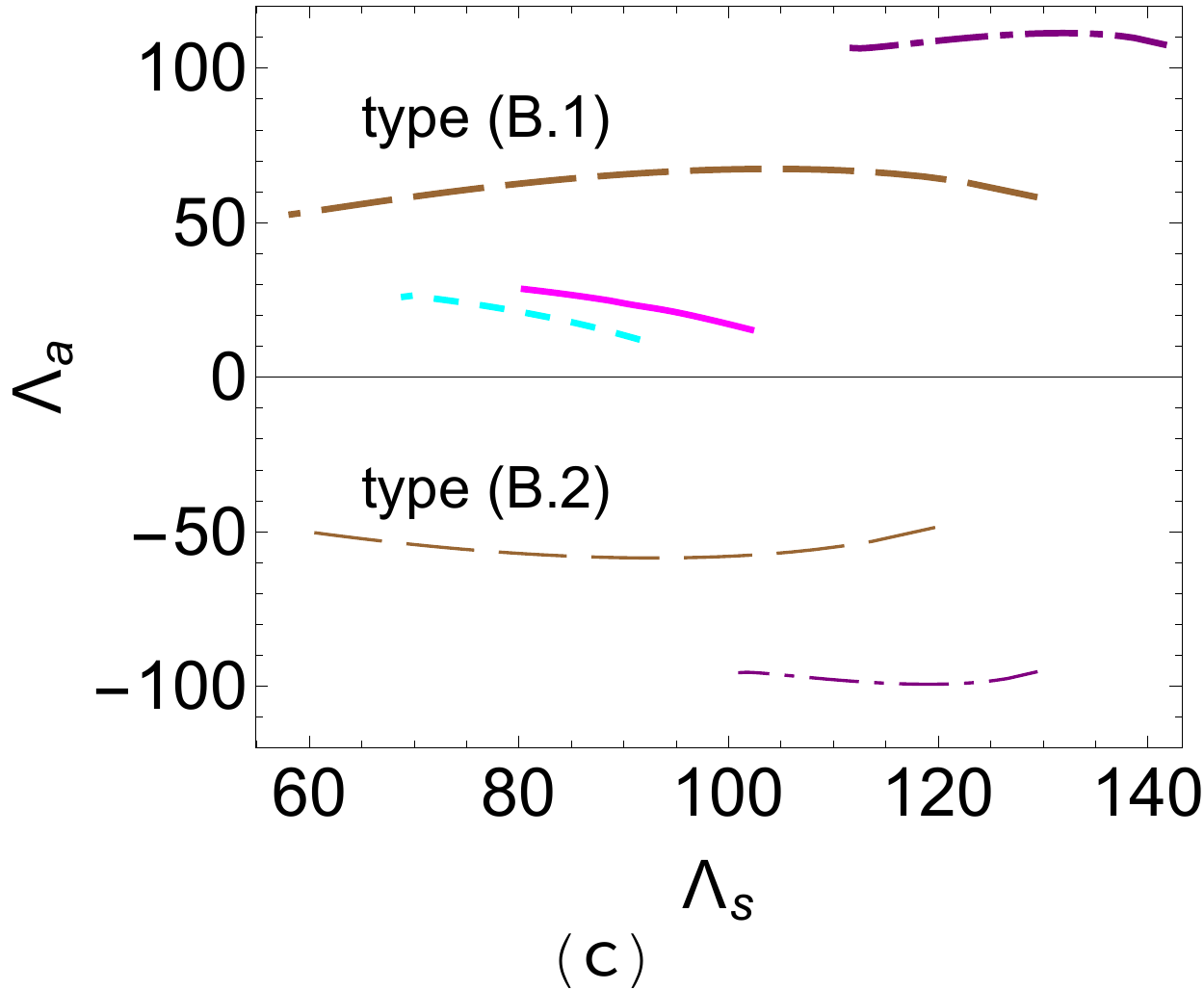} &
\includegraphics[width=0.25\linewidth]{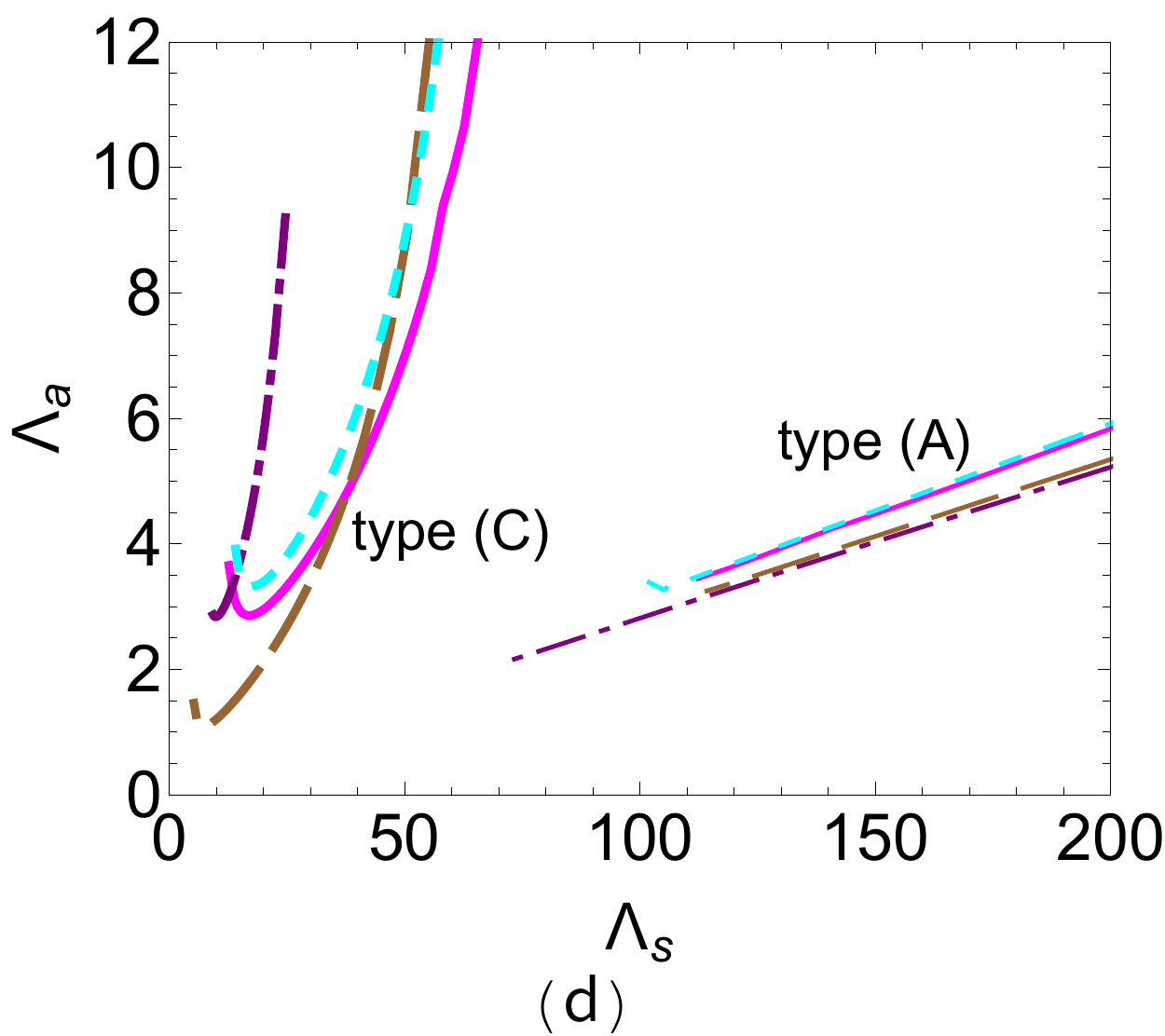} 
\end{tabular}
\caption{(Color online) Same as Fig.~\ref{fig:slope} but for EoSs that contain first-order phase transitions and a binary system with mass ratio $q=0.99$. The first-order phase transitions ($c_s=0$) of panel a introduce a second stable branch in the mass-radius curves (panel b). BLRs between stars in the same or different branches produce a slope, hill, drop, and swoosh (panels c and d). 
} 
\label{fig:swoosh}
\end{figure*}

\vspace{0.2cm}
\noindent \textit{Mass and radius with first-order phase transitions}~--~
Panel (a) in Fig.~\ref{fig:swoosh} shows $c_s^2$ for various EoSs that reproduce mass-radius curves with a  second stable branch (panel b). The first-order phase transition occurs when $c_s^2 = 0$. All EoSs are allowed by NICER's observations of J0030+0451 and J00740+6620 to 90\% confidence~\cite{Riley:2019yda,Miller:2019cac,Riley:2021pdl,Miller:2021qha}. As before, the introduction of structure at low number density changes the slope of the mass-radius curve, and allows for massive stars. This impacts both the first and second stable branches.

There are three  relevant cases in the BLRs of mass twins:
\begin{itemize}[leftmargin=*]
\setlength\parskip{0.0pt}
\setlength\parsep{0.0pt}
    \item {\bf{Type A}}.~Both stars in the first branch.
    \item {\bf{Type B}}.~One star in the first and one in the second branch.
    \item {\bf{Type C}}.~Both stars  in the second branch.
\end{itemize}
Type A binaries present BLRs similar to the above discussion, while types B and C binaries lead to different behavior.

\vspace{0.2cm}
\noindent \textit{The hill and the drop}~--~
Figure~\ref{fig:swoosh} (c) shows the BLR for type B.1 binaries with thick curves [ie.~the less massive star ($M_1$) in the first branch and the more massive one ($M_2$) in the second] and for type B.2 binaries with thin curves (vice-versa). 

For type B.2 binaries, $\Lambda_a \propto \Lambda_1 - \Lambda_2 < 0$, but for type A binaries this is not the case because the most massive star necessarily presents a lower $\Lambda$, being harder to deform. For type B.2 binaries, however, the lighter star is significantly more compact because it is in the second branch. Therefore, the lighter star is the one that is harder to deform since $\Lambda \propto C^{-p}$ with $p>0$. Some EoSs used in Fig.~\ref{fig:swoosh} for type B.1 binaries (the turquoise dashed curve and the pink solid curve) are not shown for type B.2 binaries. For type B.2 binaries to exist, we must have the more massive star in the first branch, and the less massive star in the second branch.
However, this is not practically possible for EoSs where most stars in the second branch are heavier than those in the first branch. 

The BLRs of type B.1 binaries can present either a ``drop'' (i.e.~a nearly straight line with negative slope), or  are a ``hill'' (i.e.~a nearly concave-down quadratic with a wide summit). 
The Love number always increases when mass decreases, but the Love-M relations in the first and second branch have different derivatives with respect to mass. For stars in the first branch, $|d\Lambda/dM|$ is nearly constant, but for stars in the second branch, this is not the case. For low-mass stars with certain EoSs (e.g.~EoSs 6 and 7), the magnitude of the derivative is larger in the second branch than in the first branch, but then it becomes smaller at larger masses; this is what causes the hill-like feature in the BLR. For other EoSs (EoSs 5 and 8), $|d\Lambda/dM|$ is always larger for stars in the second branch than in the first branch, leading to the drop-like feature in the BLRs. 

The inverted hill BLR of type B.2 binaries is nearly a mirror image of the hill BLR of type B.1 binaries. 
This is obviously not an exact symmetry, as one can see from the hill BLRs produced with the purple dot-dashed curve in panel (c) of Fig.~\ref{fig:swoosh}, but it good to $\lesssim  30\%$. The reason is that for both type B.1 and B.2 binaries, the Love-M relation for the star in the second branch is the same concave quadratic. Although the Love-M relation for the star in the first branch is nearly a straight line in both types of binaries, they could be shifted to lower $\Lambda_s$. The shift, however, is very small because the compactness range in the first branch of B.1 and B.2 binaries is similar, due to the small mass ratio required for type B binaries to exist.   

\vspace{0.2cm}
\noindent \textit{The swoosh}~--~
Binaries of type C have  both stars  in the second stable branch. This branch allows only a small range of masses,  forcing the mass ratio $q\sim 1$. The BLRs are EoS insensitive only when the mass ratio is away from unity \cite{Yagi:2013awa,Yagi:2013bca}. Thus, one expects the BLRs for type C stars to  not be universal. Panel (d) of Fig.~\ref{fig:swoosh} confirms this expectation and shows a curvature in the BLRs  (i.e.~non-linear structure) that resembles a ``swoosh.'' 

When both stars are in the second branch, the swoosh appears at $\Lambda_a \in (1,8)$, while when both stars are in the first branch, then the swoosh is smaller (i.e.~the curvature is smaller) and it appears only for specific EoSs. We can see this from Fig.~\ref{fig:swoosh} (d), where the swoosh is much smaller for binaries in type A than in type C. This is because for the swoosh to appear, $\Lambda_{1,2}$ has to be quite small, implying the masses of the NSs must be large,  only a property of select EoSs. The swoosh also technically appears for binaries with EoSs without first order phase transitions, provided the maximum mass is high enough, but it is an order of magnitude smaller.

All stars in the second branch are by default near the maximum mass. Therefore, an expansion in small compactness is not applicable. Instead, we must consider how the tidal deformability of either star in the binary scales with its own compactness in this high mass regime. We find numerically that $\Lambda_{1,2} \propto \mathcal{C}_{1,2}^{-9}$ for stars in the second branch (instead of $\Lambda_{1,2} \propto \mathcal{C}_{1,2}^{-5}$, which is only valid when ${\cal{C}}\ll 1$). Then  Eq.~\eqref{eq:Lratio} must be modified to $\Lambda_a/\Lambda_s = (\Lambda_1 - \Lambda_2)/(\Lambda_1 + \Lambda_2) = (\mathcal{C}_1^{-9}-\mathcal{C}_2^{-9})/(\mathcal{C}_1^{-9}+\mathcal{C}_2^{-9})$, which does not require ${\cal{C}}\ll 1$. 

Further progress requires a relation between the compactness of two stars in the second branch. Since $q\sim 1$, we can use Eq.~\eqref{eq:Taylor-C} to write $\mathcal{C}_1 = b \Lambda_2^{-1/9} q \left[1 + b \Lambda_2^{-1/9} (\Delta R/\Delta M)_1 \delta + {\cal{O}}(\delta^2)\right]$, where  $\delta = 1-q \ll 1$, $\mathcal{C}_{1,2} = b \; \Lambda_{1,2}^{-1/9}$, $\Delta M$ ($\Delta R$) the mass (radius) difference between the two stars in the binary, and $b \approx 0.35$. Taylor-expanding, 
\begin{equation}
    \Lambda_a =  \frac{9}{2} \delta \left[\Lambda_s - b \frac{\Lambda_s^{8/9}}{(\Delta M/ \Delta R)} \right] + {\cal{O}}(\delta^2)\,.
    \label{eq:swoosh-approx}
\end{equation}
This approximate BLR is controlled by the two terms inside the square bracket, with the size of $\Delta M/\Delta R$ dictating which term dominates. In the second branch, the slope $\Delta M/\Delta R$ is initially very small for the lowest mass stars (corresponding to large $\Lambda_s$). As one increases the central density (thus increasing mass and decreasing $\Lambda_s$), the slope $\Delta M/\Delta R$ reaches a maximum of ${\cal{O}}(10^{-1})$. An increase in central density beyond this decreases $\Lambda_s$ and $\Delta M/\Delta R$ further, as the maximum mass is approached. Therefore, for large values of $\Lambda_s$ (of ${\cal{O}}(10^2)$) and for small values of $\Lambda_s$ (of ${\cal{O}}(1)$), the second term dominates because $\Delta M/\Delta R$ is small. However, for intermediate values of $\Lambda_s$ (of ${\cal{O}}(10)$), both terms contribute roughly equally. This behavior is what causes the swoosh structure, and since the swoosh depends on  $\Delta M/\Delta R$, the location of the swoosh is not EoS independent, as shown in panel (d) of Fig.~\ref{fig:swoosh}. 

\vspace{0.2cm}
\noindent \textit{Implications}~--~
We have presented new physical properties of several features in the BLRs (a change in slope, a hill, a drop and a swoosh structure) that encode nuclear physics information, but can these features be extracted from GW observations? The accuracy to which the tidal deformabilities can be extracted now and in the future has been studied extensively, e.g.~in~\cite{Carson:2019rjx}. During the fifth LIGO observing run (O5), one expects to be able to measure tidal deformabilities with uncertainties of  $\delta \Lambda_{1,2} \approx  50$--$100$, while 3G detectors may allow measurements with uncertainties $\delta \Lambda_{1,2} \approx 5$--$10$. The change in slope in the BLRs may be observable already during O5 if a sufficiently loud and low mass NS binary is detected, maybe allowing for a measurement of a rise in $c_s^2$ below $3~n_{\rm sat}$.

The detection of the hill, drop or swoosh would be a smoking gun signal of mass twins, but the detectability of the latter two is more challenging. The drop and the swoosh occur at very small $\Lambda_a$, and such a detection would require uncertainties in the measurements of the tidal deformability of $\delta \Lambda_{1,2} \approx 1$ and $\approx 10$, respectively. This is beyond both 2G and 3G detectors, unless an exceptionally loud signal is detected. The hill 
structure may be detectable for sufficiently massive NSs that pass observational constraints, 
but approximate universal BLRs would not capture this behavior
Detectability would then require LIGO to consider EoSs with such structure.  The detection of the hill would decisively determine that there is a first-order phase transition inside NSs.

\vspace{0.2cm}
\noindent \textit{Acknowledgments}~--~
This work was supported in part by the National Science Foundation (NSF) within the framework of the MUSES collaboration, under grant number OAC-2103680.
J.N.H. acknowledges the support from the US-DOE Nuclear Science Grant No. DE-SC0020633. H.~T.~and N.~Y.~acknowledge support from NASA Grants No. NNX16AB98G, 80NSSC17M0041 and 80NSSC18K1352 and NSF Award No. 1759615. V.D acknowledges support from the National Science Foundation under grants PHY-1748621 and NP3M PHY-2116686, and PHAROS (COST Action CA16214). The authors also acknowledge support from the Illinois Campus Cluster, a computing resource that is operated by the Illinois Campus Cluster Program (ICCP) in conjunction with the National Center for Supercomputing Applications (NCSA), and which is supported by funds from the University of Illinois at Urbana-Champaign. 

\appendix

\bibliography{reference.bib}

\clearpage

\appendix
\onecolumngrid

\section{SUPPLEMENTARY MATERIAL}

\begin{figure}
    \centering
    \includegraphics[width=0.5\linewidth]{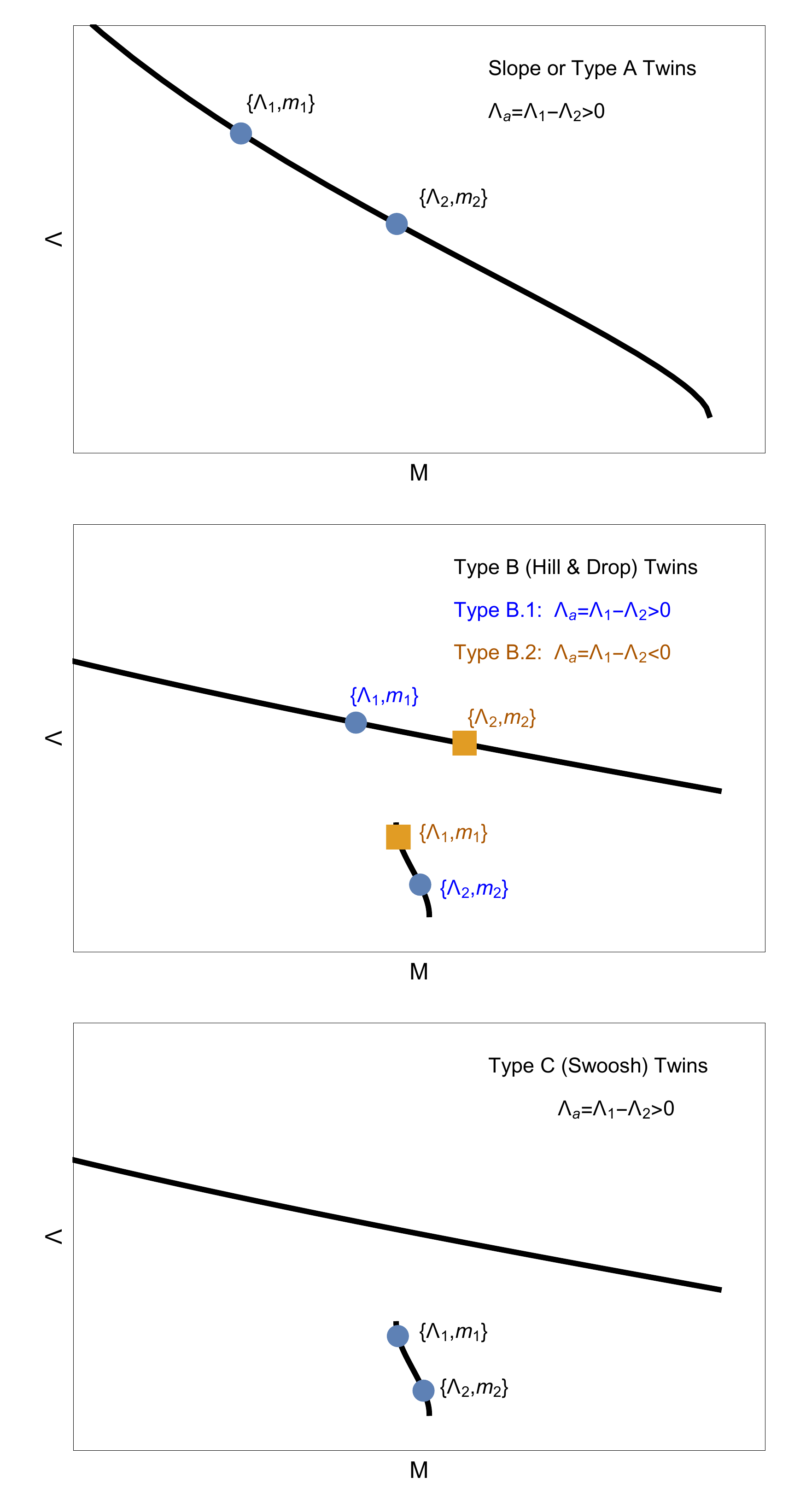}
    \caption{Diagrams showing the connection between the tidal deformability $\Lambda$, stellar mass $M$ and the type of features in $\Lambda_a$ vs. $\Lambda_s$ discussed in this work.}
    \label{fig:ski}
\end{figure}

To better understand the features of the BLRs discussed in the letter, we plot in Fig.\ \ref{fig:ski} the tidal deformability versus mass for a few relevant stellar sequences. The details of the EoSs that generate them are unimportant, except for the defining attribute of a first-order phase transition that is strong enough to produce mass twins (not required for the first case). Note that $M_1$ is \emph{always} defined as the lighter of the two stars in the binary and $M_2$ is found using a fixed mass ratio $q<1$ such that $M_2=M_1/q$. 

\begin{itemize}
    \item {\bf Slope or Type A twin stars} (top panel of Fig.\ \ref{fig:ski}): this feature arises when one compares stars found either in a single branch stellar sequence or in the first stable branch of a mass-twin sequence (which make up the vast majority of cases studied).  For these stars, $\Lambda$ decreases as $M$ increases, or $\Lambda_1>\Lambda_2$, such that $\Lambda_a>0$.  The primary distinguishing characteristic of the slope feature is \emph{how quickly} $\Lambda$ decreases as $M$ increases. 
    \item {\bf Type B twin stars} (middle panel of Fig.\ \ref{fig:ski}): 
    these features arise when one compares one star from the first stable branch with one star from the second stable branch.  There is only a limited range in q values for which this is possible. There are generally two cases to consider:
    \begin{itemize}
        \item {\bf Type B.1}: the less massive star $M_1$ is in the first stable branch (show in blue). Here $\Lambda_1>\Lambda_2$, such that $\Lambda_a>0$.  The defining feature here is that there is a discontinuity in the mass-radius sequence creating either a hill or a drop; 
        \item {\bf Type B.2}: the less massive star $M_1$ is in the second stable branch (show in yellow). Because $M_1$ is extremely compact (very large central baryon density), $\Lambda_1$ is very small, despite being a light NS.  Thus, $\Lambda_2>\Lambda_1$ and $\Lambda_a \propto \Lambda_1-\Lambda_2<0$, which uniquely generates a negative $\Lambda_a$ and an inverted hill.
    \end{itemize}
    \item {\bf Type C twin stars} (bottom panel of  Fig.\ \ref{fig:ski}): this feature arises when one compares two stars found in the second stable branch. Just like in Type A, it always holds that $\Lambda_1>\Lambda_2$, such that $\Lambda_a>0$. The fact that the second stable branch usually allows only for a small difference in mass for a large decrease in $\Lambda$ is the distinguishing characteristic that leads to a swoosh. Once more, only values of $q$ very close to 1 are possible, since most second branches are relatively short. 
\end{itemize}
 
\noindent {\bf Other considerations for Type B twins}: Some sequences of twins may only have a few stars in the first branch that possess mass twins (with a more massive second branch). In this scenario, there will be vanishingly few cases that fall into the B.2 category, making it extremely unlikely to measure a negative $\Lambda_a$ value. Thus, if a Type B.2 binary was measured (i.e. $\Lambda_a<0$), this would place tight constraints on the nature of the mass-radius sequence, and would lead to the conclusion that there is most likely a long second stable branch with mass-twins.   

\end{document}